# ~115 GeV and ~143 GeV Higgs mass considerations within the Composite Particles Model

Marko B. Popovic*

*(August 12, 2011)*

*Abstract*

The radiatively generated Higgs mass is obtained by requiring that leading "divergences" are cancelled in both 2D and 4D. This predicts one or more viable modes; the k=1 mode mass is $m_H \cong \frac{2}{3} m_t \cong 115 GeV$ whereas the k=2 mode is $m_H \cong 143 GeV$. These findings are interpreted within the Composite Particles Model (CPM), [Popovic 2002, 2010], with the massive top quark being a composite structure composed of 3 fundamental O quarks ($O\bar{O}O$) and the massive Higgs scalar being a color-neutral meson like structure composed of 2 fundamental O quarks ($\bar{O}O$). The CPM predicts that the Z mass generation is mediated primarily by a top – anti top whereas the Higgs mass is generated primarily by a O – anti O interactions. The relationship [Popovic 2010] between top Yukawa coupling and strong QCD coupling, obtained by requiring that top – anti top channel is neither attractive or repulsive at tree level at $\sqrt{s} \cong M_Z$, defines the Z mass. In addition, this relationship indirectly defines the electroweak symmetry breaking (EWSB) vacuum expectation value (VEV), the CPM Higgs mass and potentially the EWSB scale.

*Please address all correspondence to:
Marko B Popovic, Physics Department, Worcester Polytechnic Institute
100 Institute Road, Worcester, MA 01609
Cell: 617 470 8198
Email: mpopovic@wpi.edu

## 1. Introduction

The Higgs scalar particle is the last missing ingredient of the Standard Model (SM) of particle physics. The Large Electron Positron (LEP) particle accelerator near Geneva observed a number of suspicious events [1,2] in the vicinity of $115 \, GeV/c^2$, at the center of mass energies a bit above $\sqrt{s} \cong 206 \, GeV/c^2$, just before the accelerator was shut down in 2000. Recently, the Fermi National Accelerator laboratory near Chicago reported excess in dijet invariant mass [3] in the vicinity of $144 \, GeV/c^2$, at the $\sqrt{s} \cong 1.96 \, TeV/c^2$ collision energies just before the planned shutdown in September 2011.

I present a theoretical model that may be compatible with either of these two observations and it may assist the continued search at the Large Hadron Collider (LHC) at CERN near Geneva at 14 TeV energy, i.e. approximately 70,000 times larger than the scale of the strong nuclear reactions ($\approx$200 MeV). This model assumes radiative generation of the Higgs mass and cancellation of surplus leading "divergences" in both 2D and 4D within the full SM renormalization scheme applied to the running effective scalar mass squared. Two particularly interesting modes (k=1,2) are obtained and interpreted within the Composite Particles Model (CPM), [4, 5], that might closely mimic the effective SM electroweak broken phase.

According to CPM the massive top quark is a composite structure composed of 3 fundamental O quarks ($O\bar{O}O$) and the massive Higgs scalar is a color-neutral meson-like structure composed of 2 fundamental O quarks ($\bar{O}O$). Note that the SM gauge anomaly cancellation is satisfied as the elementary top quark is exchanged with the original O quark with identical SM gauge couplings.

The hierarchy problem is potentially resolved via top-anti top interactions at Z mass energy, i.e. via defining relationship [5] between top Yukawa coupling and strong coupling constant $g_{QCD}$. The three flat directions of the effective potential at Z mass energy may correspond to three long range massless bosons that couple to the SM Z boson and give the SM Z boson its mass.

## 2. Electroweak Symmetry Breaking (EWSB) within the Composite Particles Model (CPM)

The EWSB is a transition between two ground states. Here, instead of the SM Higgs and top quark fields in the unbroken massless phase the CPM [4, 5] has an auxiliary Higgs field, with zero mass and quartic coupling, i.e. $m_H = 0$ and $\lambda = 0$, and an elementary quark field, O, with Yukawa coupling equal to $1/3$ of the SM top quark Yukawa coupling. The broken massive phase is expected to closely resemble SM with the O quark field confined within the Higgs ($\bar{O}O$) and top quark ($O\bar{O}O$) composite fields.

Hence, imagine that there is no fundamental scalar field in the high-energy sector of the theory. Instead, as one of several possible scenarios, imagine that there are non-SM four-fermion interactions that may be caused by, here unspecified, broken gauge symmetries at high energies. A heavy gauge boson **G** may be integrated out and resulting contact terms may be Fiertz reordered to give the standard low-energy weak-interaction-like form of the non-SM four-fermion interaction terms.

Furthermore, imagine that product of two fermion fields within the four-fermion interaction terms may be interpreted as an effective static scalar field, $\Phi$, i.e. field without lagrangian kinetic term. As shown before [6], the static effective scalar field may generate a fully gauge covariant kinetic term in the low energy effective theory via fermion loops. This can be summarized as

$$\bar{O}GO \leftrightarrow \bar{O}GO \xrightarrow{heavy\ G\ integrated\ out} \bar{O}O\bar{O}O \xrightarrow{O\bar{O}\ identified\ as\ \Phi} \bar{O}\Phi O. \quad (1)$$

In the unbroken phase the effective lagrangian is

$$\mathcal{L}^{eff} = \mathcal{L}^{massless\ SM}_{(O\ instead\ of\ t)} + \mathcal{L}^{UV}_{(?)}, \text{ and } \langle\phi\rangle = 0. \quad (2)$$

The reminiscence of the theory in ultraviolet is expressed at low energies by an unknown lagrangian $\mathcal{L}^{UV}_{(?)}$ assumed to be irrelevant for calculations presented hereafter. The $\mathcal{L}^{massless\ SM}_{(O\ instead\ of\ t)}$ is massless SM lagrangian where top quark is exchanged with O quark with different Yukawa coupling but identical SM gauge couplings as top quark

$$\mathcal{L}^{massless\ SM}_{(O\ instead\ of\ t)} = \mathcal{L}^{SM}_{gauge\ bosons\ kinetic} + \mathcal{L}^{SM}_{fermions\ kinetic\ (t\to O)} +$$

$$+\mathcal{L}^{SM}_{Yukawa\ (t\to O)} + \begin{cases} \mathcal{L}^{SM}_{scalar\ kinetic} \\ 0 \end{cases}. \quad (3)$$

In the broken phase the effective lagrangian is identical to the SM lagrangian to extent that there may be more than one scalar field or dynamical resonances

$$\mathcal{L}^{eff} = \mathcal{L}^{massive\ SM}_{(t)} + \mathcal{L}_{massive\ scalar(s)}, \ <\phi> \neq 0. \quad (4)$$

### 2.1. Zero VEV ground state

Here, I give condition that renormalized auxiliary field stays massless in leading order in 2D and 4D.

#### 2.1.1. Mass renormalization without propagating Higgs

Leading 2D SM renormalized mass squared logarithmic "divergences", i.e. $\frac{dm_H^2(\Lambda^2)}{dln(\Lambda^2)}$ where $\Lambda$ is renormalization scale, *without* propagating Higgs, i.e. with static field, are proportional to

$$w_{2D}\frac{g_Y^2+3g_W^2}{4} - \sum\left(\frac{n_C}{3}\right)3g_f^2 \quad (5)$$

with summation over all elementary fermions and $n_C = 3\ (1)$ for quarks (leptons), see e.g. [7, 5].

In 4D, leading SM renormalized mass squared quadratic "divergences", i.e. $\frac{dm_H^2(\Lambda^2)}{d\Lambda^2}$, without propagating Higgs are proportional to

$$w_{4D}\frac{g_Y^2+3g_W^2}{4} - \sum\left(\frac{n_C}{3}\right)2g_f^2 \quad (6)$$

Factors $w_{2D} = 1$ and $w_{4D} = 2/3$ are ratios of massless over massive gauge boson polarization degrees of freedom in 2D and 4D. If couplings are identical in both 2D and 4D and "divergences" are cancelled in both 2D and 4D, a dominant Yukawa coupling squared, $g_f^2$, from Equ (5,6) equals

$$g_f^2 = \frac{g_Y^2+3g_W^2}{12}. \quad (7)$$

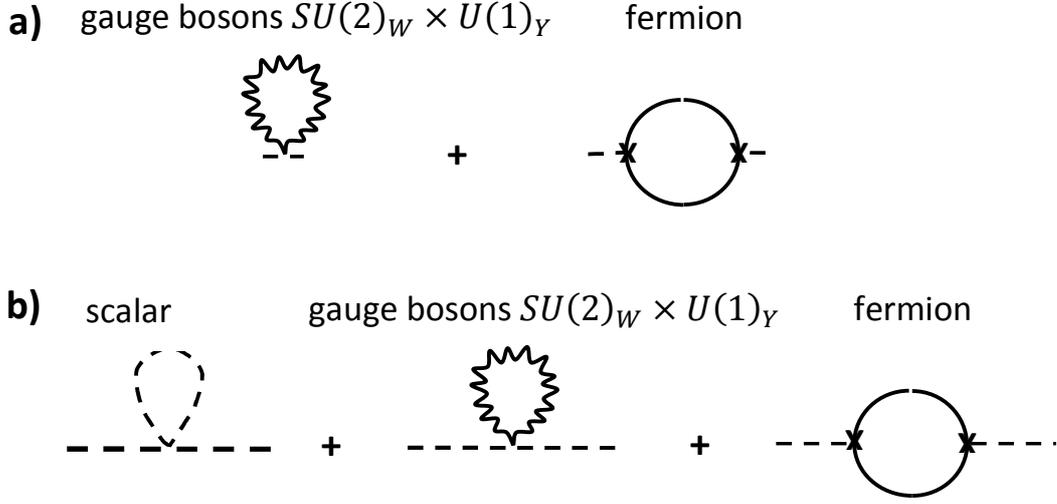

**Figure 1 a)** Auxiliary scalar field mass renormalization without scalar propagation. **b)** Auxiliary scalar field mass renormalization with propagating scalar.

### 2.1.2. Mass renormalization with propagating Higgs

In 2D, leading SM renormalized mass squared logarithmic "divergences", i.e. $\frac{dm_H^2(\Lambda^2)}{d\ln(\Lambda^2)}$, with propagating Higgs, i.e. with non-zero kinetic term, are proportional to

$$3\lambda + w_{2d}\frac{g_Y^2+3g_W^2}{4} - \sum\left(\frac{n_C}{3}\right)3g_f^2 \ . \qquad (8)$$

In 4D, leading SM renormalized mass squared quadratic "divergences", i.e. $\frac{dm_H^2(\Lambda^2)}{d\Lambda^2}$, with propagating Higgs, i.e. with non-zero kinetic term, are proportional to

$$\lambda + w_{4d}\frac{g_Y^2+3g_W^2}{4} - \sum\left(\frac{n_C}{3}\right)2g_f^2 \qquad (9)$$

From Equ (8, 9) dominant quark Yukawa coupling squared and scalar field quartic coupling equal

$$g_f^2 = \left(w_{4d} - \frac{w_{2d}}{3}\right)\frac{g_Y^2+3g_W^2}{4} \text{ and } \lambda = \left(w_{4d} - 2\frac{w_{2d}}{3}\right)\frac{g_Y^2+3g_W^2}{4} \qquad (10)$$

Hence, for expected $w_{2d} = 1$ (longitudinal) and $w_{4d} = 2/3$ (2 massless transversal over 3 massive total)

$$g_f^2 = \frac{g_Y^2+3g_W^2}{12} \text{ and } \lambda = 0. \qquad (11)$$

Therefore, the same result is obtained with and without Higgs propagation, Equ (7, 11).

### 2.1.3. Yukawa coupling defined with low energy gauge couplings

If Equ (11) or Equ (7) is satisfied for the low energy SM values of $U(1)_Y \times SU(2)_W$ couplings $g_Y, g_W$ then

$$g_f \cong \frac{g_t}{3}. \quad (12)$$

However, no SM fermion has this Yukawa coupling and SM Higgs quartic coupling is not zero.

### 2.2. Non-zero VEV ground state

I address the condition that radiatively generated scalar mass is in agreement with renormalization group equations in 2D description. Why is 2D description important? As is demonstrated by the lattice arguments, e.g. see [8], the non-Abelian gauge fields carry charge that causes their propagation to mimic the 1-space dimensional flux providing confinement between static charges. Hence, the 2D considerations here are thought of as consequence of non-Abelian gauge fields' dynamics in regular 4D.

As first emphasized by Nambu, in simplest models with dynamical mass generation [9-11] and top condensate [12-15] EWSB, one may expect the Higgs mass on the order of 2 top quark masses. Hence, if top quark is exchanged with O quark one might expect the Higgs mass on the order of $\frac{2}{3} m_t$ [4, 5].

#### 2.2.1. Transition at low energies: 2D consideration

Let us consider EWSB, where "original" fields and parameters, abruptly change across phase transition and Higgs mass is $\frac{2}{3} m_t$. The effect of "transition" can be expressed by an unknown parameter A

$$g_f \cong \frac{g_t}{3} \to A g_f \cong A \frac{g_t}{3}, \quad (13a)$$

$$m_H = 0 \to m_H = \frac{2}{3} m_t \text{ or } \lambda = 0 \to \lambda = \frac{m_H^2}{v_{EW}^2} = \frac{4 m_t^2}{9 v_{EW}^2} = \frac{2 g_t^2}{9}. \quad (13b)$$

Imposing the cancelation of leading "divergences" in 2D, Equ (8), with help of Equ (7, 11,13b) leads to

$$3\lambda + \frac{g_Y^2 + 3 g_W^2}{4} = \frac{6 g_t^2}{9} + 3 g_t^2 \left(\frac{1}{3}\right)^2 = 3 g_t^2 \left(\frac{1}{3}\right)^2 A^2 \quad (14)$$

$$\Rightarrow A^2 = 3 \text{ or } g_f \cong \frac{g_t}{3} \to \sqrt{3} g_f \cong \frac{g_t}{\sqrt{3}}. \quad (15)$$

However, not all "divergences" should cancel because Higgs mass is expected to be radiatively generated. Hence, the physical top Yukawa coupling is $g_t \cong 1$ whereas $\frac{g_t}{\sqrt{3}} \cong \frac{1}{\sqrt{3}}$ is the factor defining the fermion loop contribution to cancelation of just a part of leading "divergences". I explain that next.

#### 2.2.2. Effect of radiatively generated Higgs mass: 2D calculation and scalar mass predictions

I split dominant fermion loop into two terms; the first term (x-term) contributes to the cancellation of leading "divergences" in Equ (8) and the second term (y-term) equals to the radiatively generated Higgs mass. Two terms add to one, i.e. $x + y = 1$ and the single color 2D fermion loop is exactly solved [5] as

$$2D \; fermion \; loop = \frac{k g^2}{\pi}. \quad (16)$$

This result is obtained with similar techniques as the fermion loop in the Schwinger model [16]. However, in contrast to the Schwinger model this is a composite scalar and not a gauge boson. Hence, if "fermion loop" is indeed identified as a "*single* fermion loop" in a pure 2D calculation (i.e. no 4D spin) one would expect k=1. However, to make the connection with 4D, one should also consider a factor of k=2 here, due to the scalar nature of the propagator, as fermion spins may point inward or outward (when 4D spin is also considered). Hence, I leave an explicit dependence on the relevant phase space parameterized with k. Finally, in 2D, according to Schwinger model, the mass singularities in the propagator should exist for multiple integer values [8, 16], i.e. $k \in N$. The "*multi*-fermion loop" interpretation of $k = 1, 2$ modes and reason for $g = g_t$ are provided in 2.2.3.

The obtained system has three unknowns, $x, y$ and $\lambda$, and three equations,

$$3\lambda + \frac{g_Y^2 + 3g_W^2}{4} = 3g_t^2 x \;, \quad x + y = 1 \quad \text{and} \quad \lambda = \frac{kg_t^2}{\pi} y, \tag{17}$$

leading to an unique solution

$$3\lambda + \frac{g_Y^2 + 3g_W^2}{4} = 3g_t^2 \left(1 - \frac{\lambda \pi}{kg_t^2}\right) \rightarrow \sqrt{\lambda} = \sqrt{\frac{g_t^2 - \frac{g_Y^2 + 3g_W^2}{12}}{1 + \frac{\pi}{k}}}, \tag{18}$$

$$\Rightarrow m_H = \sqrt{\frac{6m_t^2 - M_Z^2 - 2M_W^2}{3\left(1 + \frac{\pi}{k}\right)}}. \tag{19}$$

The above calculation, see [5], is self consistent as the Higgs mass in the Higgs loop propagator (within piece proportional to $x$) is identical to radiatively generated Higgs mass (within piece proportional to $y$).

For the world average top quark mass, $m_t = 173.1 \pm 1.3 \; GeV$, I obtain

$$m_H = \begin{cases} 113.0 \pm 1.0 \; GeV \; for \; k = 1 \rightarrow y = \mathbf{0.669}, \; x = 0.331 \\ 143.4 \pm 1.3 \; GeV \; for \; k = 2 \rightarrow y = \mathbf{0.539}, \; x = 0.461 \end{cases}. \tag{20}$$

Hence, for the k=1 mode, the parameterized effective top Yukawa coupling, specific to cancellation of part of the "divergences" is equal to $g_t\sqrt{0.331} \cong \frac{1}{\sqrt{3}}$, exactly as anticipated with Equ (15)! And predicted Higgs mass is roughly as originally anticipated, i.e. $m_H \cong \frac{2}{3}m_t = 115.4 \; GeV$!

Interestingly, according to Equ (19), for the $k \rightarrow \infty$ the $m_H \rightarrow \cong 230 \; GeV$. This finding is compatible with the upper limit on the SM Higgs mass [20, 5] obtained from the requirement that the running effective Higgs mass takes zero value at order $\sim TeV$ scale. The mapping between the physical SM Higgs mass and this scale, named Higgs Mass Zero Crossing (HMZC) scale, has been presented in [20, 5].

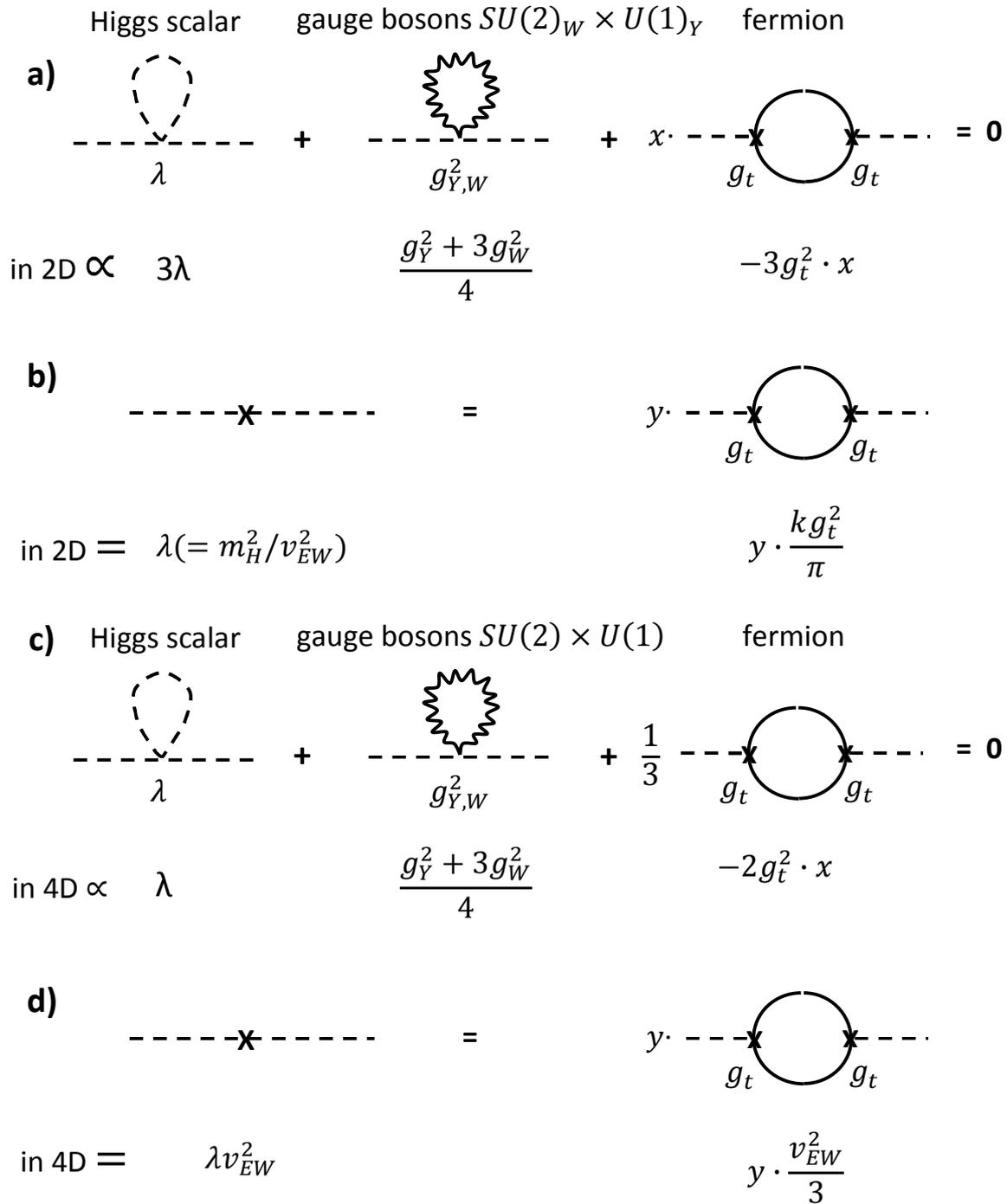

**Figure 2 a)** Self-consistent cancelation of "divergences" and **b)** radiative generation of Higgs mass in 2D. **c)** Self-consistent cancelation of "divergences" in 4D and **d)** radiative generation of Higgs mass in 4D.

### 2.2.3. Interpretation of 2D k=1 mode from O-anti O condensate

The Higgs mass squared in Equ (17) within 2D k=1 mode may be rewritten as

$$y \frac{g_t^2}{\pi} \cong \frac{2}{3} \frac{g_t^2}{\pi} = 6 \frac{(g_t/3)^2}{\pi} = n_S n_C \frac{(g_t/3)^2}{\pi} \tag{21}$$

where $n_S = 2$ and $n_C = 3$ are spin and color contributions to phase space respectively and where $\cong g_t/3$ is O Yukawa coupling. Hence, 2D k=1 Higgs appears to be generated via O -anti O condensate.

### 2.2.4. Interpretation of 2D k=2 mode from O-anti O and top-anti top condensates mixing

The Higgs mass squared in Equ (17) within 2D k=2 mode may be rewritten as

$$y \frac{2g_t^2}{\pi} \cong 0.539 \frac{2g_t^2}{\pi} = 0.539 \cdot 2 \cdot 9 \frac{(g_t/3)^2}{\pi} = 9.702 \frac{(g_t/3)^2}{\pi}. \tag{22}$$

Now, I hypothesize that top – anti-top condensate within CPM may also contribute to 2D k=2 Higgs as

$$n_S n_C \pi^2 \frac{(g_t/3)^2}{\pi} \tag{23}$$

where factor $\pi^2$ is the phase space of 2 transversal O – anti O condensates with axial $(0, \pi)$ symmetry.

Here, I expect that weighted sum of 2D k=1 mode and contribution from the top –anti top condensate should be equal to 2D k=2 Higgs mass and parameterized with an angle $\alpha$, which may be calculated as

$$\cos^2 \alpha \cdot n_S n_C + \sin^2 \alpha \cdot n_S n_C \pi^2 \cong 9.702 \quad \Rightarrow \quad \cos^2 \alpha = 0.930 \tag{24}$$

### 2.2.5. 4D matching

Next, I address k=1 mode in 4D. Again the "fermion loop" is split into x and y components

$$\lambda + \frac{g_Y^2 + 3g_W^2}{4} - 2g_t^2 x = 0 \tag{25}$$

and the radiatively generated Higgs mass is product of y, 4D single color fermion loop, and $\cos^2 \alpha$,

$$\lambda v_{EW}^2 = y \cdot \text{4D fermion loop} \cdot \cos^2 \alpha. \tag{26}$$

If each color contributes the same amount to VEV squared, i.e. $v_{EW}^2 = 3 \cdot (\text{4D fermion loop})$, then

$$\lambda = y \cdot \frac{1}{3} \cdot \cos^2 \alpha = (1 - x) \cdot \frac{1}{3} \cdot \cos^2 \alpha \quad \Rightarrow \quad x = 1 - \frac{3}{\cos^2 \alpha} \lambda. \tag{27}$$

Substituting expression for x back into Equ (25), for angle $\alpha$ as in Equ (24), gives

$$\sqrt{\lambda} = \sqrt{\frac{2g_t^2 - \frac{g_Y^2 + 3g_W^2}{4}}{1 + \frac{6}{\cos^2 \alpha} g_t^2}} \quad \Rightarrow \quad m_H = \sqrt{\frac{4m_t^2 - M_Z^2 - 2M_W^2}{1 + \frac{12}{\cos^2 \alpha} \frac{m_t^2}{v_{EW}^2}}} \cong 115.6 \, GeV. \tag{28}$$

with 4D $x = 0.289, y = 0.711$ and scalar mass differs only 0.2 $GeV$ from the CPM k=1 prediction.

However, if 2D k=1 mode's $x = 0.311$ is enforced in 4D then from Equ (25)

$$\lambda + \frac{g_Y^2 + 3g_W^2}{4} - 2g_t^2 x = 0 \Rightarrow m_H = \sqrt{4 \cdot x \cdot m_t^2 - M_Z^2 - 2M_W^2} = 138.1 \, GeV. \quad (29)$$

This result was obtained by Popovic [4, 5] and it may be compared with a "long lived" SM solution [4, 5] where both dimensionless parameters of the Higgs potential tend towards zero at ~ the Planck scale.

By adding top – anti top contribution to 4D O – anti O result, Equ (29), I obtain

$$m_H = 138.1 \, GeV \cdot \sqrt{1 + \tan^2 \alpha} = 143.2 \, GeV \quad (30)$$

with the same angle as in Equ (24, 26); the scalar mass differs only 0.2 $GeV$ from the CPM k=2 prediction.

It would be worthwhile investigating whether the Fermi lab findings [2], i.e. Gaussian excess centered at 144 GeV dijet invariant mass, originate indirectly from (1) processes involving W boson, CPM O and bottom quarks (sum of their masses is $M_W + \frac{m_t}{3} + m_b \cong 142.5 \, GeV$) or directly from (2) decay of k=2 mode 143.4GeV Higgs interacting with both O –anti O (96.4%) and top –anti top (3.6%) condensates.

### 3. The Bose-Einstein distribution applied to condensates mixing

Higgs field, $\Phi$, may be shared by O -anti O, top – anti top condensate, and maybe something else

$$\Phi = c_2 \, O\bar{O} + c_6 \, t\bar{t} + \cdots \quad (31)$$

where coefficients $c_2, c_6, \cdots$ (subscripts are chosen to remind that this is CPM model) are relative contributions normalized to one. By applying the Bose-Einstein distribution with assumed values

$$\hbar\omega_2 \cong m_{o\bar{o}} c^2 \cong kT \, , \quad \hbar\omega_6 \cong m_{t\bar{t}} c^2 \cong 3 \, m_{o\bar{o}} c^2 \, , \quad (32)$$

$$\Rightarrow c_2 = \frac{\frac{1}{e^1 - 1}}{\frac{1}{e^1 - 1} + \frac{1}{e^3 - 1} + \cdots} = \frac{0.58197}{0.58197 + 0.05239 + \cdots} \leq 0.91741 \quad (33)$$

with $m_{t\bar{t}} = 2m_t$ and $m_{O\bar{O}} = \frac{2}{3} m_t$ on the lines of top condensate models as emphasized by Nambu [15]. The mass contribution to Higgs field from each condensate type can be expressed as

$$m_H = c_2 \cdot 2 \frac{m_t}{3} + c_6 \cdot 6 \frac{m_t}{3} + \cdots \quad (34)$$

$$\Rightarrow m_H \geq 0.91741 \cdot 2 \frac{m_t}{3} + (1 - 0.91741) 6 \cdot \frac{m_t}{3} \text{ or } m_H \geq 134.5 \, GeV. \quad (35)$$

Clearly this result is dependent on $kT$. For $kT \cong M_Z c^2$ I obtain $c_2 \leq 0.94477$ and $m_H \geq 128.1 \, GeV$. However, for the pole-consistent solutions, $kT \cong m_H c^2$, I obtain the smallest possible Higgs mass as

$$c_2 \leq 0.89484 \text{ and } m_H \geq 139.7 \, GeV, \quad (36)$$

in a close agreement with k=2 Higgs mass addressed above. This result again corresponds to

$$\cos^2 \alpha \cong [1 - a(1 - c_2)]^2 = 0.931 \text{ for } a = \frac{1}{3} = \frac{m_{o\bar{o}}}{m_{t\bar{t}}}. \tag{37}$$

### *4. Z mass and condition for the top – anti top long range massless mode*

Here, I investigate whether the SM $\bar{t}t$ channel at energy of Z pole mass is repulsive or attractive.

Consider the $\bar{t}t$ scattering in the Euclidean space and ignore chiralities of the incoming and outgoing particles while assuming that left and right handed tops are equally represented within particle and antiparticle solution. The main interaction channels at tree level are gluon and Higgs exchange. The weak interactions are absent as interacting particles have opposite chiralities and the hypercharge interactions are zero due to the equal sharing conjecture introduced above.

I now assume that strong QCD interactions proportional to $-g_{QCD}^2 \, T_{aij}T_{akl}$, where $a = 1 \ldots 8$, $i,j = 1,2,3$ and summation over repeated indices is implied, are exactly *balanced* with the Yukawa forces due to the virtual Higgs particle exchange proportional to $g_t^2$ as a condition for the loose bound state, see Fig 3.

$$\frac{1}{3}g_s^2 \times 2 \qquad -g_t^2 \quad = \mathbf{0} \rightarrow \alpha_s = \frac{3}{2}\alpha_t = \frac{3g_t^2}{8\pi} = 0.1181 \pm 0.0018$$

$$\text{World average } \alpha_s \cong 0.1184 \pm 0.0007$$

**Figure 3** Finely balanced interplay between the QCD gluon and Higgs scalar mediated top anti-top interactions.

Hence, the back of the envelope calculation at tree level suggests

$$2\frac{1}{2}\frac{2}{3}g_{QCD}^2 = g_t^2 \quad or \quad \alpha_S = \frac{3}{2}\frac{g_t^2}{4\pi} \quad \text{at Z mass} \tag{38}$$

where I used $(T^a)_{ij}(T^a)_{kl} = \frac{1}{2}\left(\delta_{il}\delta_{kj} - \frac{1}{N}\delta_{ij}\delta_{kl}\right)$ for $SU(N)$ groups, see for example [17], where $N = 3$. For the *colorless* composite Higgs the expression in bracket equals $\frac{2}{3}$, and additional factor of 2 in Equ (38) corresponds to two transversal gluon polarizations. **QED**

The result in Equ (38) is in an excellent agreement with the standard estimate of the strong running coupling constant [18, 19]. Equ (38) predicts $\alpha_s = 0.1181 \pm 0.0018$ given the world average top quark mass $m_t = 173.1 \pm 1.3\ GeV$ where uncertainty is therefore solely due to the top quark mass uncertainty. This can be compared with the current world average value $\alpha_s \cong 0.1184 \pm 0.0007$ at $s = M_Z^2$ [18, 19].

Even if the equal distribution assumption is ignored and hypercharge interactions are taken into account that would change the above result only on the order $\frac{3}{2} Y_L Y_R \frac{g_Y^2}{g_t^2} = \frac{3}{2}\frac{1}{6}\frac{2}{3}\frac{g_Y^2}{g_t^2} \sim 2\%$ .

Interpretation of this result is that it takes zero energy to orient the top – anti top at Z scale; these excitations are the long range massless bosons (i.e. flat directions of effective potential) that couple to the massless SM Z boson and give Z mass. That is not the case with original field O as $g_{QCD}^2 \gg \frac{3}{2}\left(\frac{g_t}{3}\right)^2$. The tops' interactions are: 1) zero at Z scale, 2) attractive at larger distances (QCD coupling grows toward smaller energies) and 3) repulsive at small distances. Hence Equ (38) is likely the underlying principle defining the EWSB VEV, the CPM scalar mass(es) and potentially the EWSB scale, spanning vast energies in a natural fashion and removing the hierarchy problem.

## 5. Summary

I review earlier CPM results [4, 5] and introduce the concept of condensates mixing. The CPM predicts the Z mass generation mediated primarily by top – anti top interactions whereas the physical Higgs mass is generated primarily by O – anti O interactions. The radiatively generated k=1 mode Higgs mass is expected to be $m_H \cong \frac{2}{3} m_t \cong 115\ GeV$ whereas k=2 mode mass is $m_H \cong 143\ GeV$. These predictions are supported by self-consistent cancellation of extra leading "divergences" in both 2D and 4D.

The hierarchy problem is resolved by defining relationship Equ (38), [5]. This relationship between top Yukawa coupling and strong QCD coupling obtained by requiring that top – anti top channel is neither attractive or repulsive at tree level at $\sqrt{s} \cong M_Z$, defines the Z mass and hence indirectly the EW VEV and the CPM Higgs mass. Finally, Equ (38) defines the HMZC scale, [20, 5], at which the running effective Higgs mass squared is zero. As argued in [5] the HMZC$\cong$EWSB scale if SM is the effective theory in vicinity of the EWSB scale, and if the Universe dynamics was never dominantly tachyonic.

**Acknowledgement**

I am indebted to Gorazd Cvetic, Thomas Hambye, Dejan Stojkovic, and Chris T. Hill, for valuable comments and encouragements to publish earlier and unfortunately much longer version of this manuscript. I am thankful to everyone at the WPI Physics Department for giving me a warm welcome and physics home last year. I am indebted to Kristin Politi for editing part of this manuscript.